\documentclass[11pt,aps,preprintnumbers,amsmath,amssymb,superscriptaddress,nofootinbib]{revtex4}
\usepackage{epsfig,subfigure}
\usepackage{epstopdf}
\usepackage[utf8]{inputenc}
\usepackage{graphicx}
\usepackage{amssymb}
\usepackage{amsxtra}
\usepackage{amsmath}
\usepackage{booktabs,multirow,tabularx}
\usepackage{slashed}
\usepackage{float}
\usepackage{placeins}
\usepackage{rotating}
\usepackage{lscape}
\usepackage{color}
\usepackage{hyperref}
\usepackage{soul}
\usepackage{subfigure}

\newcommand{\bea}{\begin{eqnarray}}
\newcommand{\eea}{\end{eqnarray}}

\def\beq{\begin{equation}}
\def\eeq{\end{equation}}
\newcommand{\vev}[1]{\langle {#1} \rangle}

\def\gev{\rm GeV}

\begin{document}
\title{Complex Scalar Singlet Model Benchmarks for Snowmass}

\author{Shekhar Adhikari}
\email{sadhika8@jhmi.edu}
\affiliation{Department of Physics and Astronomy, University of Kansas, Lawrence, Kansas, 66045~ U.S.A.}
\affiliation{Department of Radiation Oncology and Molecular Radiation Sciences, School of Medicine,
Johns Hopkins University, Baltimore, Maryland 21231}

\author{Samuel D. Lane}
\email{samuel.lane@ku.edu}
\affiliation{Department of Physics and Astronomy, University of Kansas, Lawrence, Kansas, 66045~ U.S.A.}

\author{Ian M. Lewis}
\email{ian.lewis@ku.edu}
\affiliation{Department of Physics and Astronomy, University of Kansas, Lawrence, Kansas, 66045~ U.S.A.}

\author{Matthew Sullivan}
\email{msullivan1@bnl.gov}
\affiliation{Department of Physics and Astronomy, University of Kansas, Lawrence, Kansas, 66045~ U.S.A.}
\affiliation{High Energy Theory Group, Physics Department, Brookhaven National Laboratory,
Upton, New York, 11973~ U.S.A.}

\begin{abstract}
{\bf Executive Summary:} In this contribution to Snowmass 2021, we present benchmark parameters for the general  complex scalar singlet model. The complex scalar singlet extension has three massive scalar states with interesting decay chains which will depend on the exact mass hierarchy of the system. We find maximum branching ratios for resonant double Standard Model-like Higgs production, resonant production of a Standard Model-like Higgs and a new scalar, and double resonant new scalar production.  These branching ratios are between 0.7 and 1.  This is particularly interesting because instead of direct production, the main production of a new scalar resonance may be from the $s$-channel production and decay of another scalar resonance.  That is, it is still possible for discovery of new scalar resonances to be from the cascade of one resonance to another.  We choose our benchmark points to have to have a large range of signatures: multi-$b$ production, multi-$W$ and $Z$ production, and multi-125 GeV SM-like Higgs production.  These benchmark points can provide various spectacular signatures that are consistent with current experimental and theoretical bounds.  This is a summary of results in Ref.~\cite{ToAppear}.

\end{abstract}
\maketitle

\section{Introduction}
\label{sec:intro}
As the search for new physics continues, the high luminosity Large Hadron Collider (HL-LHC)  could very well provide the first evidence of beyond the Standard Model (BSM) physics.  One of the simplest BSM scenarios is the addition of new real or complex scalar states that are singlets under the Standard Model (SM) gauge group. These complex scalar singlets also appear in more complete models~\cite{Muhlleitner:2017dkd,Abouabid:2021yvw}, and can help in solving fundamental questions in the field such as being dark matter candidates~\cite{Gonderinger:2012rd,Coimbra:2013qq,Muhlleitner:2020wwk}. These simple singlet extensions have been extensively studied under the assumption they have some additional softly broken symmetries such as a $U(1)$ or $\mathbb{Z}_2$~\cite{Costa:2015llh,Muhlleitner:2020wwk}. Complex scalar singlet extensions are particularly interesting because there are two scalar states in addition to the Higgs boson.  Indeed, it could be that both new resonances could be discovered by one decaying into the other.

In this paper we summarize results from Ref.~\cite{ToAppear}.  We consider the general complex scalar singlet extension of the SM with no additional symmetries~\cite{Dawson:2017jja}.  This model extends the SM by two new CP even scalars. We find benchmark points that maximize the various di-scalar resonant productions at the HL-LHC: double 125 GeV SM-like Higgs bosons, SM-like Higgs in association with a new scalar, and two heavy new scalar bosons.  This model is equivalent to the SM extended by adding two real scalar singlet extension with no additional symmetries beyond the SM.  Benchmarks for two real singlet extensions with $Z_2$ symmetries have been studied previously~\cite{Robens:2019kga,Papaefstathiou:2020lyp}.  In section~\ref{sec:model}, we introduce the model and discuss the phenomenology of the scalar sector. In section~\ref{sec:constraints} we explore the current constraints on the model and in section~\ref{sec:benchmarks} present various benchmark points of phenomenological interest for the High Luminosity upgrade at the Large Hadron Collider (HL-LHC).

\section{Model}
\label{sec:model}
Following Ref.~\cite{Dawson:2017jja}, we use the most general scalar potential involving the complex scalar singlet, $S_c=(S_0+i\,A)/\sqrt{2}$,  and the Higgs doublet, $\Phi=(0,(v_{EW}+h)/\sqrt{2})^{\rm T}$ in the unitary gauge. $S_0$, $A$, and $h$ are all real CP even scalar fields, and $v_{EW}=246$~GeV is the Higgs vacuum expectation value. The scalar potential can be written as 
\begin{eqnarray}
V(\Phi,S_c)&=&\frac{\mu^2}{2}\Phi^\dagger \Phi+\frac{\lambda}{4}(\Phi^\dagger\Phi)^4+\frac{b_2}{2}|S_c|^2+\frac{d_2}{4}|S_c|^4+\frac{\delta_2}{2}\Phi^\dagger\Phi |S_c|^2\nonumber\\
&&+\left(a_1\,S_c+\frac{b_1}{4}\,S_c^2+\frac{e_1}{6}\,S_c^3+\frac{e_2}{6}S_c|S_c|^2+\frac{\delta_1}{4} \Phi^\dagger\Phi\,S_c+\frac{\delta_3}{4}\Phi^\dagger\Phi\,S_c^2\right.\nonumber\\
&&\left.+\frac{d_1}{8}S_c^4+\frac{d_3}{8}S_c^2|S_c|^2+{\rm h.c.}\right)~\label{eq:VSc}
\end{eqnarray}
where $a_1,b_1,e_1,e_2,\delta_1,\delta_3,d_1,d_3$ are complex parameters. As shown in Refs.~\cite{Chen:2014ask,Lewis:2017dme,Dawson:2017jja}, we can set $\langle S_c\rangle =0$ without loss of generality.  

The model contains three scalar mass eigenstates, $h_1$, $h_2$ and $h_3$ with masses $m_1$, $m_2$, and $m_3$, respectively. We will take $h_1$ to be the discovered Higgs boson with mass $m_1 = 125~\gev $. The mass eigenstates can be obtained from the gauge states via a $SO(3)$ rotation with three rotation angles, $\theta_1, \theta_2$, and $\theta_3$.  The $\theta_3$ angle may be removed by appropriate choice of $S_c$ phase~\cite{Dawson:2017jja}. Taking the small mixing limit in $\theta_2$, the mass eigenstates are given by transformation
\begin{eqnarray}
\begin{pmatrix} h_1\\h_2\\h_3\end{pmatrix} =\begin{pmatrix} \cos\theta_1 & -\sin\theta_1 & 0\\\sin\theta_1 & \cos\theta_1 & \sin\theta_2 \\ \sin\theta_1\sin\theta_2 & \cos\theta_1 \sin\theta_2 & - 1\end{pmatrix}\begin{pmatrix} h\\S_0\\A\end{pmatrix}+\mathcal{O}(\sin^2\theta_2).\label{eq:mixinglim}
\end{eqnarray}
The couplings of $h_2$ and $h_3$ to SM fermions and gauge bosons are inherited via the mixing with the SM-like Higgs boson.  We see that $h_2$ will couple to SM fermions and gauge bosons with couplings suppressed by a factor of $\sin \theta_1$, regardless of the size of $\theta_2$. Thus, we expect $h_2$ productions modes will be similar to that of the SM Higgs but with mass of $m_2$. 

The coupling of $h_3$ to SM fermions and gauge bosons is doubly suppressed by the factor $\sin\theta_1 \sin \theta_2$. Therefore, we expect the dominant production of $h_3$ to be from decays of $h_2$, when it is kinematically allowed. With this in mind, we will restrict ourselves to to the mass ordering $m_2>m_3>m_1$.

\section{Constraints}
\label{sec:constraints}
The theoretical constraints we consider are narrow width, perturbative unitarity, boundedness, and global minimization. We restrict our parameters such that the total width of $h_2$ is less than $10\%$ of its mass. 
We ensure perturbative unitarity is not violated at tree level by first computing the $J=0$ partial wave matrix for two-to-two scalar scattering through the quartic couplings. Then we numerically diagonalize and make sure the eigenvalues are less than $1/2$.  Finally we check that the numerically found global minima of the potential corresponds to the electroweak minima, $\vev{\Phi} = (0,v_{EW}/\sqrt{2})^{\rm T}$ and $\vev{S_c} = 0$, where $v_{EW}=246$~GeV.

We now turn to the current experimental constraints on the model. Note that all SM-like rates and branching ratios are taken from the LHC Higgs Cross Section Working group suggested values~\cite{deFlorian:2016spz}. First, we consider the signal strengths of Higgs precision measurements. In our model the production cross sections for $h_1$ are suppressed by a factor of $\cos^2 \theta_1$, while the branching ratios remain unchanged. Thus we expect for each production mode $i$ and decay chain $i\rightarrow h_1 \rightarrow f$ the signal strength is
\begin{eqnarray}
\mu_i^f=\frac{\sigma_i(pp\rightarrow h_1){\rm BR}(h_1\rightarrow f)}{\sigma_{i,\rm SM}(pp\rightarrow h_1){\rm BR}_{\rm SM}(h_1\rightarrow f)}=\cos^2\theta_1,
\end{eqnarray}
where the subscript ${\rm SM}$ indicates SM values, and the numerator is calculated in the complex scalar singlet model.  
We then fit the mixing angle $\theta_1$ using a $\chi^2$ fit to the measured signal strengths~\cite{ToAppear}.

Next, we turn our attention to the direct searches for heavy scalars \cite{ToAppear}. We will need the production cross section and branching ratios to SM final states in order to implement these constraints. As stated in section ~\ref{sec:model} the couplings between $h_2$ and fermions and gauge bosons are suppressed by a factor of $\sin\theta_1$. Thus, the production rates and partial widths are given by 
\begin{eqnarray}
\sigma(pp\rightarrow h_2)\approx \sin^2\theta_1 \sigma_{\rm SM}(pp\rightarrow h_2),\quad \Gamma(h_2\rightarrow f_{\rm SM})\approx \sin^2\theta_1 \Gamma_{\rm SM}(h_2\rightarrow f_{\rm SM}),
\end{eqnarray}
where $\sigma_{\rm SM}$ and $\Gamma_{\rm SM}$ indicate SM Higgs rates at the mass $m_2$ and $f_{\rm SM}$ are SM gauge bosons and fermions. We also consider the decay widths for $h_2 \rightarrow$  $h_1h_1$, $h_1h_3$, or $h_3h_3$, when the masses place us in the kinematically allowed region. 

Normally, a ``hard cut'' is imposed to determine such constraints. Parameter points are rejected if their predicted cross sections are greater than any observed limit. However, this does not allow for large fluctuations for individual channels with small fluctuations in other channels.  On the other hand if we use our method detailed in \cite{Adhikari:2020vqo}, we construct a channel-by-channel $\chi^2$ for the heavy resonant searches to consistently combine all heavy scalar search channels and the Higgs signal strength measurements. In this method the $\chi^2$ squared function for each channel is
\begin{eqnarray}
\left(\chi^{f}_{i,h_2}\right)^2=\begin{cases}
\displaystyle\left(\frac{\sigma_i(pp\rightarrow h_2){\rm BR}(h_2\rightarrow f)+\hat{\sigma}_{i,Exp}^f-\hat{\sigma}_{i,Obs}^f}{\hat{\sigma}_{i,Exp}^f/1.96}\right)^2 & {\rm if~} \hat{\sigma}_{i,Obs}^f\geq \hat{\sigma}_{i,Exp}^f\\
\displaystyle\left(\frac{\sigma_i(pp\rightarrow h_2){\rm BR}(h_2\rightarrow f)}{\hat{\sigma}_{i,Obs}^f/1.96}\right)^2&{\rm if~} \hat{\sigma}_{i,Obs}^f< \hat{\sigma}_{i,Exp}^f.
\end{cases}~\label{eq:Chi2Scalar}
\end{eqnarray}
where $\sigma_i(pp\rightarrow h_2)$ is the resonance production cross section from initial state $i$, ${\rm BR}(h_2\rightarrow f)$ is the branching ratio into final state $f$, $\hat{\sigma}_{i,Exp}^f$ ($\hat{\sigma}_{i,Obs}^f$) is the experimentally determined expected (observed) 95\% CL upper limit on $\sigma(i\rightarrow h_2){\rm BR}(h_2\rightarrow f)$.  For a single channel, this reproduces the traditional ``hard cut'' method, but allows us to combine multiple channels into a global $\Delta \chi^2$.

\begin{figure}[tb]
\begin{center}
    \subfigure[]{\includegraphics[width=0.45\textwidth,clip]{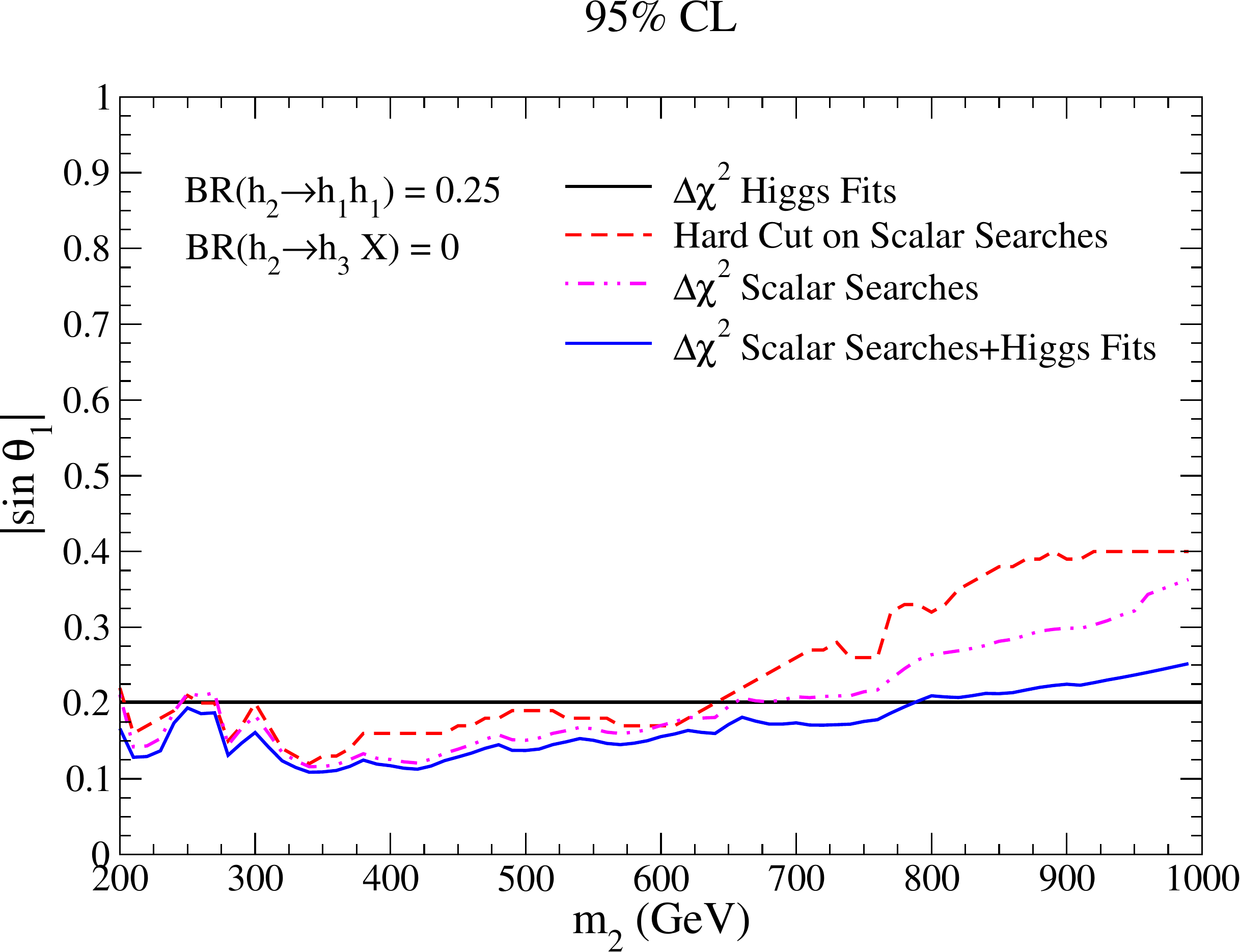}}
    \subfigure[]{\includegraphics[width=0.45\textwidth,clip]{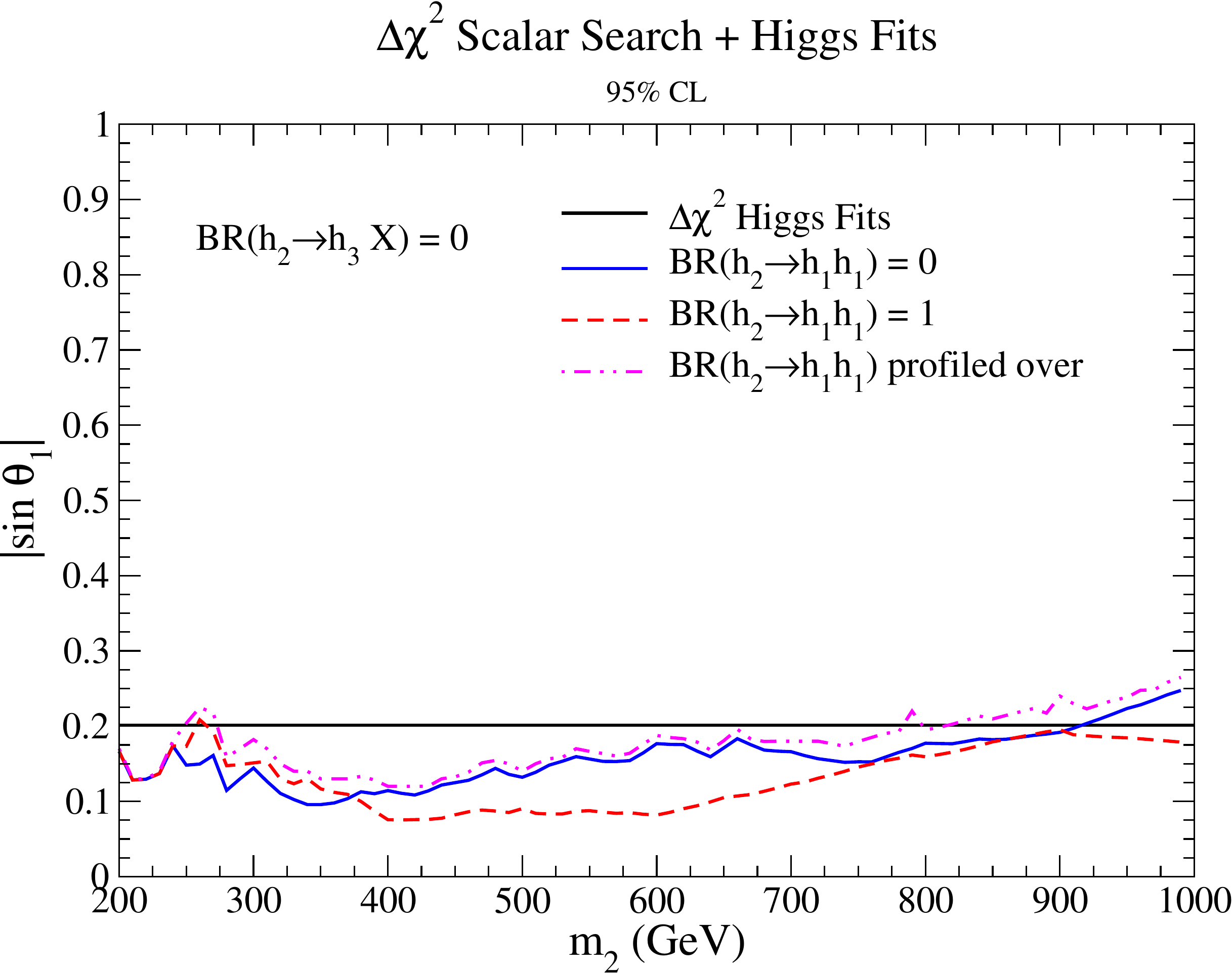}}
\end{center}
\caption{\label{fig:sth}  In both (a) and (b) black solid lines show $\Delta\chi^2$ fits to Higgs signal strength data. (a) Bounds on $\sin \theta_1$ with ${\rm BR}(h_2 \rightarrow h_1 h_1)$=0.25 for (red dashed) ``hard cuts'' on scalar resonance searches, (magenta dot-dot-dashed) $\Delta\chi^2$ fit to scalar resonance searches, and (blue solid) combined $\Delta \chi^2$ fits to Higgs precision and resonant scalar searches. (b) Comparison of combined $\Delta\chi^2$ fits to Higgs precision data and resonant scalar searches for (blue solid) ${\rm BR}(h_2\rightarrow h_1h_1)=0$, (red dashed) ${\rm BR}(h_2\rightarrow h_1 h_1)=1$, and (magenta dot-dot-dashed) profiling over ${\rm BR}(h_2\rightarrow h_1 h_1)$.  In both (a,b) ${\rm BR}(h_2\rightarrow h_1 h_3)={\rm BR}(h_2\rightarrow h_3 h_3)=0$.}
\end{figure}

In Figure~\ref{fig:sth}(a) we compare the resulting $95\%$ confidence level constraints on $| \sin\theta_1 |$ vs $m_2$ using a Higgs signal strength fit (solid black), heavy scalar searches using a traditional hard cut (dashed red), heavy scalar searches fitting a combined $\Delta \chi^2$ [Eq.~(\ref{eq:Chi2Scalar})] across relevant channels (dot-dot-dashed magenta), and the total combined $\Delta \chi^2 $ for heavy scalar searches and Higgs fits (solid blue). We have taken BR$(h_2 \rightarrow h_3 X) = 0$ for $X = h_1$ or $h_3$. This will correspond to the most constraining case since this will force $h_2$ to decay to only SM final states. Here we see that for the heavy scalar searches that the $\Delta \chi^2$ are consistently stronger than the traditional hard cut.  However, for $m_2\gtrsim 650$~GeV, Higgs signal strengths are stronger than the hard cuts.  Hence, in the usual method the Higgs signal strength bound $|\sin\theta_1|\lesssim 0.2$ would be used.  However,  for $m_2\gtrsim 800$~GeV, the combined $\Delta \chi^2$ is less constraining than the Higgs signal strength fits since our method allows for more fluctuation.  

In Figure~\ref{fig:sth}(b), we show the comparison of $95\%$ confidence level constraints on $| \sin\theta_1 |$ vs $m_2$ using the $\Delta \chi^2$ method for Higgs Fits (solid black) and Higgs signal strength fits + direct scalar searches for BR$(h_2 \rightarrow h_1 h_1) = 0,1,$ and profiled (respectively solid blue, dashed red, and dot-dot-dashed magenta). We see that profiling ${\rm BR}(h_2\rightarrow h_1h_1)$ is the least constraining, while the most constraining alternates between BR$(h_2 \rightarrow h_1 h_1) = 0$ and $1$. We will take the most constraining $\sin \theta_1$ from this plot for our benchmark points.  

\section{Benchmark Points}
\label{sec:benchmarks}
\begin{figure}[tb]
\begin{center}
    \subfigure[]{\includegraphics[width=0.45\textwidth,clip]{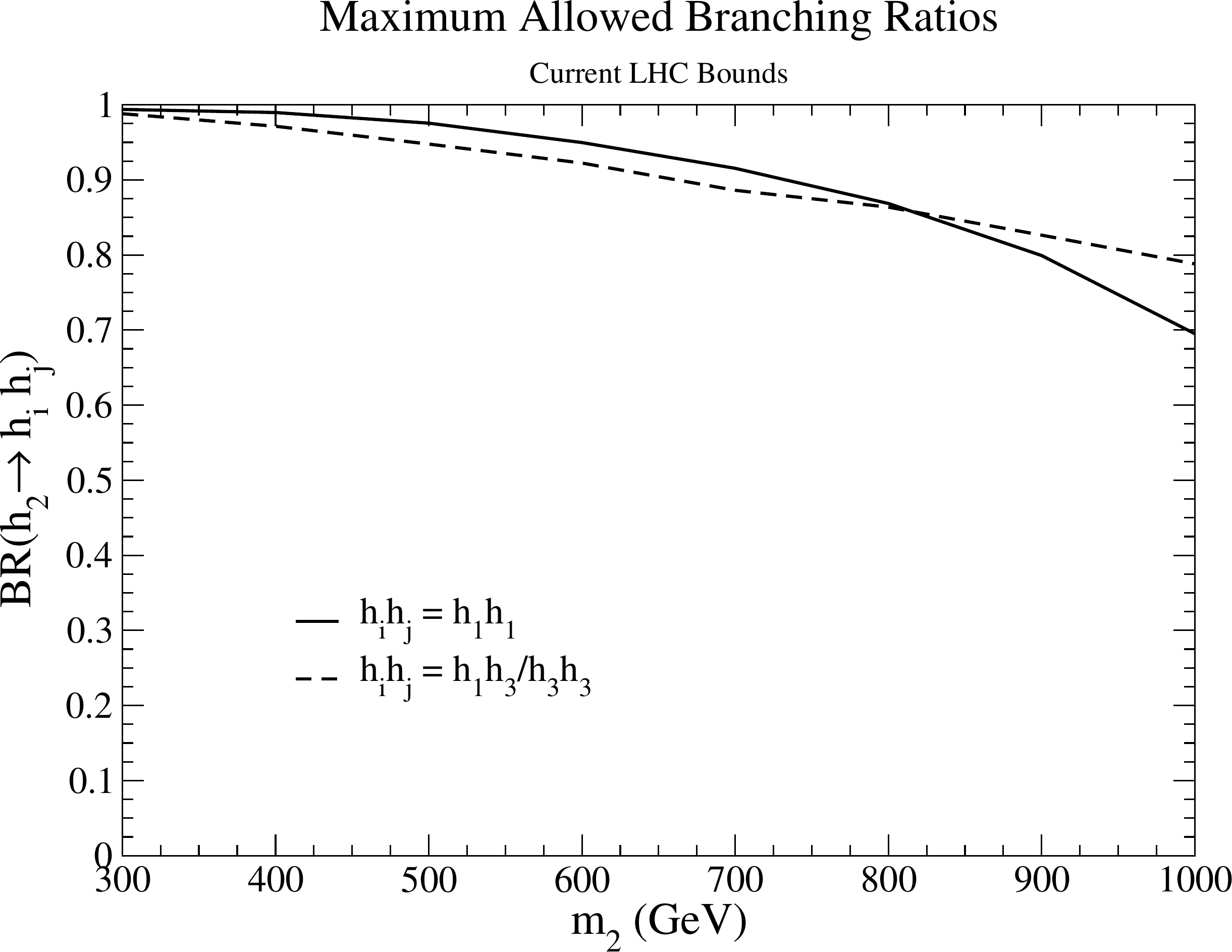}}\\
    \subfigure[]{\includegraphics[width=0.45\textwidth,clip]{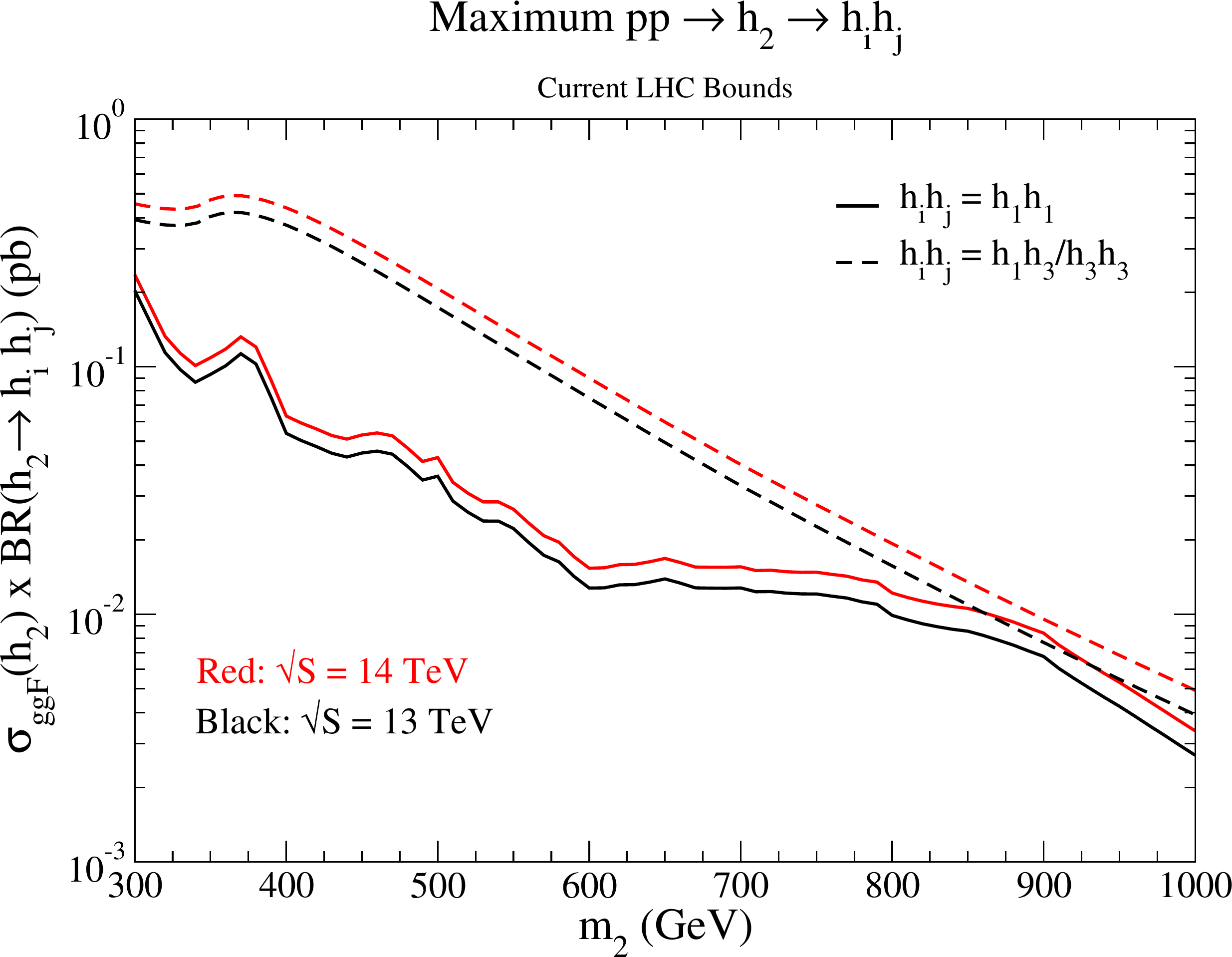}}
    \subfigure[]{\includegraphics[width=0.45\textwidth,clip]{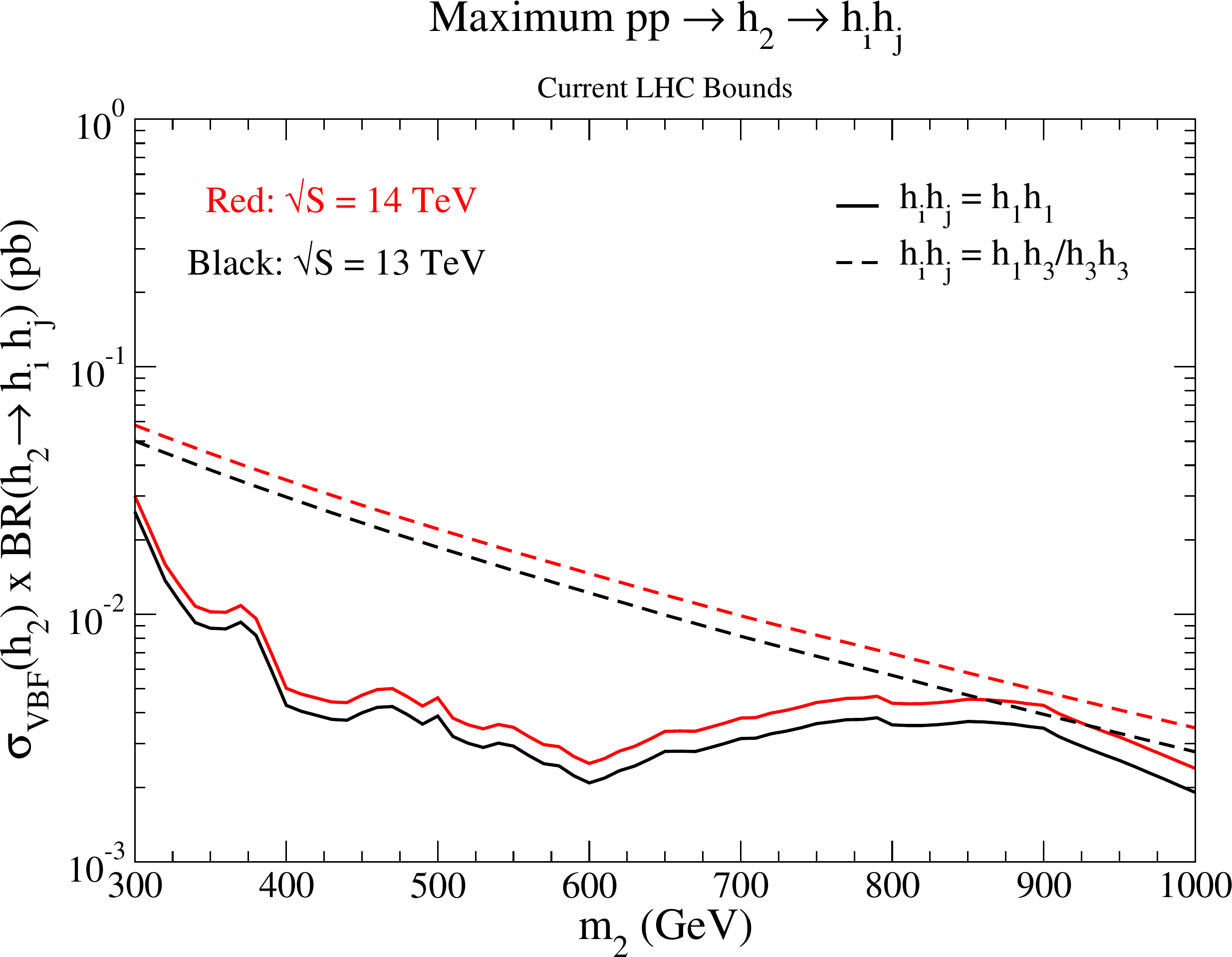}}
\end{center}
\caption{\label{fig:benchmarks} (a) Maximum allowed branching ratios with current LHC data for (solid) $h_1h_1$ resonance and (dashed) $h_1h_3$ and $h_3h_3$ resonance.  (c,d) Maximum $h_2$ production and decay rates for (solid) $h_2\rightarrow h_1h_1$ and (dashed) $h_2\rightarrow h_1h_3/h_3h_3$.  Red lines are for a 14 TeV LHC and black for a 13 TeV LHC.  Both (c) gluon fusion and (d) vector boson fusion production rates are shown.  It is required that $\Gamma_{\rm Tot}(h_2)\leq 0.1\,m_2$.}
\end{figure}

Our benchmarks are created by maximizing resonant di-scalar production while keeping the total width of $h_2$ less than 10$\%$ of $m_2$.  In practice, for current $\sin\theta_1$ bounds, this means maximizing the branching ratios of a resonant scalar $h_2$ into double SM-like Higgs bosons $h_2\rightarrow h_1h_1$, a SM-like Higgs boson and new scalar $h_2\rightarrow h_1h_3$, and two new scalars $h_2\rightarrow h_3h_3$. The maximum ${\rm BR}(h_2\rightarrow h_1h_3)$ and ${\rm BR}(h_2\rightarrow h_3h_3)$ will be large enough to effectively nullify direct heavy scalar search bounds.  Hence, for $h_2\rightarrow h_1h_3$ and $h_2\rightarrow h_3h_3$ we only consider $\sin\theta_1$ constraints from precision Higgs signal strength measurements and set $\sin\theta_1=0.201$.  For $h_2\rightarrow h_1h_1$ direct scalar searches are relevant.  Hence, conservatively, we set $\sin\theta_1$ to be the minimum of all constraints in Fig.~\ref{fig:sth}(b).

 The results are shown in Fig.~\ref{fig:benchmarks} for (a) maximum branching ratios, (b) maximum $h_2$ production and decay rates in the gluon fusion channel, and (c) maximum $h_2$ production and decay rates in the vector boson fusion channel.  Some comments are in order:
\begin{itemize}
    \item The maximum branching ratios of $h_2\rightarrow h_1h_3$ and $h_2\rightarrow h_3h_3$ are the same.  Additionally, while kinematically allowed, the maximum branching ratios are independent of the mass of $h_3$.  (We have checked this for $m_3=130,\,200,$ and $270$ GeV, as shown in Tabs.~\ref{tab:bm213}~\ref{tab:bm233}).  This can be understood by noting that for a given total width $\Gamma_{\rm Tot}(h_2)$, $h_2$ branching ratios have an upper limit\\
    \begin{equation}
    {\rm BR}(h_2\rightarrow h_i h_j)\leq 1-\frac{\sin^2\theta_1 \Gamma_{\rm SM}(h_2)}{\Gamma_{\rm Tot(h_2)}},\label{eq:max}
    \end{equation}
    where $\Gamma_{\rm SM}(h_2)$ is the total width of a SM-like Higgs with mass $m_2$. There is enough freedom in this model such that maximum branching ratios for $h_2\rightarrow h_1h_3$ and $h_2\rightarrow h_3h_3$ in Fig.~\ref{fig:benchmarks}(a) saturate this bound for $\Gamma_{\rm Tot}(h_2)=0.1\,m_2$.
    \item The maximum $h_2\rightarrow h_1h_1$ is different than $h_2\rightarrow h_1h_3$ and $h_2\rightarrow h_3h_3$.  First, this is because the $\sin\theta_1$ used is different.  As we showed in Ref.~\cite{Lewis:2017dme}, for smaller mixing angles we can get large branching ratios.  Although, as shown in Fig.~\ref{fig:benchmarks}(b,c) the rates are smaller.  
    
    The other effect is that $h_2\rightarrow h_1h_1$ does not always saturate the maximum in Eq.~(\ref{eq:max}).  In the small angle limit, the relevant scalar trilinear couplings are
    \begin{eqnarray}
    h_1 h_1 h_2&:&\sin\theta_1\frac{m_2^2+2\,m_1^2-[{\rm Re}(\delta_3)+\delta_2]\,v^2}{v}+\mathcal{O}(\sin^2\theta_1),\nonumber\\
    h_1 h_2 h_3&:&\frac{{\rm Im}(\delta_3)}{2}v+\mathcal{O}(\sin\theta_1),\\
    h_2h_3h_3&:& -\frac{1}{\sqrt{2}}\left({\rm Re}(e_1)-\frac{1}{3}{\rm Re}(e_2)\right)+\mathcal{O}(\sin\theta_1).\nonumber
    \end{eqnarray}
    The $h_2-h_1-h_1$ coupling has the same $\sin\theta_1$ suppression as the couplings of $h_2$ to SM gauge bosons and fermions.  Hence, for $h_2\rightarrow h_1h_1$ to saturate the maximum branching ratio bound, the quartics ${\rm Re}(\delta_3)$ and $\delta_2$ have to be very large.  However, perturbative unitarity bounds place strong constraints on this couplings.
\end{itemize}

\begin{table}[tb]
\begin{center}
\makebox[0.9\textwidth]{
    \begin{tabular}{|c|c|c|ll|l|}\hline\hline
        $m_2$& $m_3$ & BRs and width & \multicolumn{2}{c|}{$\sigma(pp\rightarrow h_2\rightarrow h_1h_1)$} & Parameters  \\\hline\hline
       \multirow{12}{*}{400 GeV} &\multirow{4}{*}{130 GeV} & &13 TeV ggF:& $54$ fb & $d_2=0.190,\,\delta_2=23.1,\,\delta_3=22.7+i\,0.0000871$\\
		& & ${\rm BR}(h_2\rightarrow h_1h_1)=0.99$& 13 TeV VBF:& $4.3$ fb &$d_1=-0.132-i\,0.00764,\,d_3=0.0485-i\,0.000618$\\
		& &$\Gamma_{\rm Tot}(h_2)=0.041\,m_2$& 14 TeV ggF:& $63$ fb &$e_1=(-33.3-i\,14.7)v,\,e_2=(-99.6+i\,46.5)v$\\
		& &  & 14 TeV VBF:& $5.0$ fb &$\sin\theta_1=0.0756$\\ \cline{2-6}
          &\multirow{4}{*}{200 GeV} & &13 TeV ggF:& $54$ fb & $d_2=0.22,\,\delta_2=25.2,\,\delta_3=24.2+i\,0.0914$\\
		& & ${\rm BR}(h_2\rightarrow h_1h_1)=0.99$& 13 TeV VBF:& $4.3$ fb &$d_1=-0.211-i\,0.00610,\,d_3=-0.00157+i\,0.0000325$\\
		& &$\Gamma_{\rm Tot}(h_2)=0.046\,m_2$ & 14 TeV ggF:& $63$ fb &$e_1=(-29.1-i\,11.7)v,\,e_2=(-92.6+i\,36.9)v$\\
		& &  & 14 TeV VBF:& $5.0$ fb &$\sin\theta_1=0.0756$\\ \cline{2-6}
        &\multirow{4}{*}{270 GeV} & &13 TeV ggF:& $54$ fb & $d_2=0.22,\,\delta_2=25.2,\,\delta_3=24.2+i\,0.0914$\\
		& & ${\rm BR}(h_2\rightarrow h_1h_1)=0.99$& 13 TeV VBF:& $4.3$ fb &$d_1=-0.211-i\,0.00610,\,d_3=-0.00157+i\,0.0000325$\\
		& &$\Gamma_{\rm Tot}(h_2)=0.046\,m_2$& 14 TeV ggF:& $63$ fb &$e_1=(-29.1-i\,11.7)v,\,e_2=(-92.6+i\,36.9)v$\\
		& &  & 14 TeV VBF:& $5.0$ fb &$\sin\theta_1=0.0756$\\ \cline{1-6}
        \multirow{12}{*}{600 GeV} &\multirow{4}{*}{130 GeV} & &13 TeV ggF:& $13$ fb & $d_2=0.869,\,\delta_2=24.2,\,\delta_3=23.9+i\,0.0243$\\
		& & ${\rm BR}(h_2\rightarrow h_1h_1)=0.95$& 13 TeV VBF:& $2.1$ fb &$d_1=-0.356+i\,0.122,\,d_3=-0.343-i\,0.0415$\\
		& &$\Gamma_{\rm Tot}(h_2)=0.026\,m_2$& 14 TeV ggF:& $15$ fb &$e_1=(-33.2-i\,10.8)v,\,e_2=(-99.4+i\,31.9)v$\\
		& &  & 14 TeV VBF:& $2.5$ fb &$\sin\theta_1=0.0819$\\ \cline{2-6}
          &\multirow{4}{*}{200 GeV} & &13 TeV ggF:& $13$ fb & $d_2=0.869,\,\delta_2=24.2,\,\delta_3=23.9+i\,0.0243$\\
		& & ${\rm BR}(h_2\rightarrow h_1h_1)=0.95$& 13 TeV VBF:& $2.1$ fb &$d_1=-0.356+i\,0.122,\,d_3=-0.343-i\,0.0415$\\
		& &$\Gamma_{\rm Tot}(h_2)=0.026\,m_2$& 14 TeV ggF:& $15$ fb &$e_1=(-33.2-i\,10.8)v,\,e_2=(-99.4+i\,31.9)v$\\
		& &  & 14 TeV VBF:& $2.5$ fb &$\sin\theta_1=0.0819$\\ \cline{2-6}
        &\multirow{4}{*}{270 GeV} & &13 TeV ggF:& $13$ fb & $d_2=0.869,\,\delta_2=24.2,\,\delta_3=23.9+i\,0.0243$\\
		& & ${\rm BR}(h_2\rightarrow h_1h_1)=0.95$& 13 TeV VBF:& $2.1$ fb &$d_1=-0.356+i\,0.122,\,d_3=-0.343-i\,0.0415$\\
		& &$\Gamma_{\rm Tot}(h_2)=0.026\,m_2$& 14 TeV ggF:& $15$ fb &$e_1=(-33.2-i\,10.8)v,\,e_2=(-99.4+i\,31.9)v$\\
		& &  & 14 TeV VBF:& $2.5$ fb &$\sin\theta_1=0.0819$\\ \cline{1-6}
        \multirow{12}{*}{800 GeV} &\multirow{4}{*}{130 GeV} & &13 TeV ggF:& $9.9$ fb & $d_2=0.611,\,\delta_2=24.6,\,\delta_3=23.5+i\,0.00901$\\
		& & ${\rm BR}(h_2\rightarrow h_1h_1)=0.87$& 13 TeV VBF:& $3.6$ fb &$d_1=-0.0806+i\,0.368,\,d_3=-0.128-i\,0.0143$\\
		& &$\Gamma_{\rm Tot}(h_2)=0.066\,m_2$& 14 TeV ggF:& $12$ fb &$e_1=(-33.0+i\,28.5)v,\,e_2=(-99.4-i\,91.9)v$\\
		& &  & 14 TeV VBF:& $4.4$ fb &$\sin\theta_1=0.159$\\ \cline{2-6}
          &\multirow{4}{*}{200 GeV} & &13 TeV ggF:& $9.9$ fb & $d_2=0.611,\,\delta_2=24.6,\,\delta_3=23.5+i\,0.00901$\\
		& & ${\rm BR}(h_2\rightarrow h_1h_1)=0.87$& 13 TeV VBF:& $3.6$ fb &$d_1=-0.0806+i\,0.368,\,d_3=-0.128-i\,0.0143$\\
		& &$\Gamma_{\rm Tot}(h_2)=0.066\,m_2$& 14 TeV ggF:& $12$ fb &$e_1=(-33.0+i\,28.5)v,\,e_2=(-99.4-i\,91.9)v$\\
		& &  & 14 TeV VBF:& $4.4$ fb &$\sin\theta_1=0.159$\\ \cline{2-6}
        &\multirow{4}{*}{270 GeV} & &13 TeV ggF:& $9.9$ fb & $d_2=0.611,\,\delta_2=24.6,\,\delta_3=23.5+i\,0.00901$\\
		& & ${\rm BR}(h_2\rightarrow h_1h_1)=0.87$& 13 TeV VBF:& $3.6$ fb &$d_1=-0.0806+i\,0.368,\,d_3=-0.128-i\,0.0143$\\
		& &$\Gamma_{\rm Tot}(h_2)=0.066\,m_2$& 14 TeV ggF:& $12$ fb &$e_1=(-33.0+i\,28.5)v,\,e_2=(-99.4-i\,91.9)v$\\
		& &  & 14 TeV VBF:& $4.4$ fb &$\sin\theta_1=0.159$\\ \cline{1-6}
    \end{tabular}}
\end{center}
\caption{\label{tab:bm211} Benchmark points that maximize ${\rm BR}(h_2\rightarrow h_1h_1)$ with cross sections at the LHC.}
\end{table}

\begin{table}[tb]
\begin{center}
\makebox[0.9\textwidth]{
    \begin{tabular}{|c|c|c|ll|l|}\hline\hline
        $m_2$& $m_3$ & BRs and width & \multicolumn{2}{c|}{$\sigma(pp\rightarrow h_2\rightarrow h_1h_3)$} & Parameters  \\\hline\hline
       \multirow{12}{*}{400 GeV} &\multirow{4}{*}{130 GeV} & &13 TeV ggF:& $370$ fb & $d_2=22.9,\,\delta_2=3.18,\,\delta_3=-0.332+i\,0$\\
		& & ${\rm BR}(h_2\rightarrow h_1h_3)=0.97$& 13 TeV VBF:& $30$ fb &$d_1=-4.86-i\,3.37,\,d_3=-3.88-i\,2.68$\\
		& &$\Gamma_{\rm Tot}(h_2)=0.1\,m_2$& 14 TeV ggF:& $440$ fb &$e_1=(-0.250-i\,61.0)v,\,e_2=(-2.28+i\,94.9)v$\\
		& &  & 14 TeV VBF:& $35$ fb &$\sin\theta_1=0.201$\\ \cline{2-6}
          &\multirow{4}{*}{200 GeV} & &13 TeV ggF:& $370$ fb & $d_2=18.5,\,\delta_2=1.25,\,\delta_3=-0.0573+i\,0$\\
		& & ${\rm BR}(h_2\rightarrow h_1h_3)=0.97$& 13 TeV VBF:& $30$ fb &$d_1=-5.71-i\,2.78,\,d_3=-7.49-i\,8.61$\\
		& &$\Gamma_{\rm Tot}(h_2)=0.1\,m_2$& 14 TeV ggF:& $440$ fb &$e_1=(7.65+i\,39.5)v,\,e_2=(-21.4-i\,16.4)v$\\
		& &  & 14 TeV VBF:& $35$ fb &$\sin\theta_1=0.201$\\ \cline{2-6}
        &\multirow{4}{*}{270 GeV} & &13 TeV ggF:& $370$ fb & $d_2=18.7,\,\delta_2=0.197,\,\delta_3=-0.0000418+i\,0.134$\\
		& & ${\rm BR}(h_2\rightarrow h_1h_3)=0.97$& 13 TeV VBF:& $30$ fb &$d_1=7.83+i\,2.51,\,d_3=0.493+i\,3.96$\\
		& &$\Gamma_{\rm Tot}(h_2)=0.1\,m_2$& 14 TeV ggF:& $440$ fb &$e_1=(72.0+i\,86.0)v,\,e_2=(-92.5-i\,54.7)v$\\
		& &  & 14 TeV VBF:& $35$ fb &$\sin\theta_1=0.201$\\ \cline{1-6}
        \multirow{12}{*}{600 GeV} &\multirow{4}{*}{130 GeV} & &13 TeV ggF:& $75$ fb & $d_2=18.2,\,\delta_2=3.41,\,\delta_3=0.258+i\,0$\\
		& & ${\rm BR}(h_2\rightarrow h_1h_3)=0.92$& 13 TeV VBF:& $12$ fb &$d_1=5.97+i\,2.24,\,d_3=2.38+i\,7.29$\\
		& &$\Gamma_{\rm Tot}(h_2)=0.1\,m_2$& 14 TeV ggF:& $90$ fb &$e_1=(-4.59+i\,37.6)v,\,e_2=(-15.1+i\,6.20)v$\\
		& &  & 14 TeV VBF:& $15$ fb &$\sin\theta_1=0.201$\\ \cline{2-6}
          &\multirow{4}{*}{200 GeV} & &13 TeV ggF:& $75$ fb & $d_2=20.8,\,\delta_2=1.72,\,\delta_3=0.503+i\,0$\\
		& & ${\rm BR}(h_2\rightarrow h_1h_3)=0.92$& 13 TeV VBF:& $12$ fb &$d_1=6.25+i\,1.80,\,d_3=-4.63+i\,6.12$\\
		& &$\Gamma_{\rm Tot}(h_2)=0.1\,m_2$& 14 TeV ggF:& $90$ fb &$e_1=(-7.24+i\,59.1)v,\,e_2=(-22.2-i\,53.3)v$\\
		& &  & 14 TeV VBF:& $15$ fb &$\sin\theta_1=0.201$\\ \cline{2-6}
        &\multirow{4}{*}{270 GeV} & &13 TeV ggF:& $75$ fb & $d_2=17.9,\,\delta_2=0.467,\,\delta_3=-0.0976+i\,0.0946$\\
		& & ${\rm BR}(h_2\rightarrow h_1h_3)=0.92$& 13 TeV VBF:& $12$ fb &$d_1=4.16-i\,2.35,\,d_3=3.27-i\,3.49$\\
		& &$\Gamma_{\rm Tot}(h_2)=0.1\,m_2$& 14 TeV ggF:& $90$ fb &$e_1=(-11.8+i\,57.7)v,\,e_2=(-35.7-i\,39.9)v$\\
		& &  & 14 TeV VBF:& $15$ fb &$\sin\theta_1=0.201$\\ \cline{1-6}
        \multirow{12}{*}{800 GeV} &\multirow{4}{*}{130 GeV} & &13 TeV ggF:& $16$ fb & $d_2=19.9,\,\delta_2=3.22,\,\delta_3=2.98+i\,0$\\
		& & ${\rm BR}(h_2\rightarrow h_1h_3)=0.86$& 13 TeV VBF:& $5.7$ fb &$d_1=6.44-i\,0.319,\,d_3=3.90-i\,1.23$\\
		& &$\Gamma_{\rm Tot}(h_2)=0.1\,m_2$& 14 TeV ggF:& $19$ fb &$e_1=(-8.89-i\,61.0)v,\,e_2=(-26.8+i\,33.1)v$\\
		& &  & 14 TeV VBF:& $6.9$ fb &$\sin\theta_1=0.201$\\ \cline{2-6}
          &\multirow{4}{*}{200 GeV} & &13 TeV ggF:& $16$ fb & $d_2=21.1,\,\delta_2=4.54,\,\delta_3=1.76+i\,0.605$\\
		& & ${\rm BR}(h_2\rightarrow h_1h_3)=0.86$& 13 TeV VBF:& $5.7$ fb &$d_1=6.74+i\,2.11,\,d_3=3.07-i\,10.1$\\
		& &$\Gamma_{\rm Tot}(h_2)=0.1\,m_2$& 14 TeV ggF:& $19$ fb &$e_1=(-11.8-i\,46.7)v,\,e_2=(-36.8-i\,6.65)v$\\
		& &  & 14 TeV VBF:& $6.9$ fb &$\sin\theta_1=0.201$\\ \cline{2-6}
        &\multirow{4}{*}{270 GeV} & &13 TeV ggF:& $16$ fb & $d_2=18.9,\,\delta_2=4.20,\,\delta_3=2.06-i\,0.137$\\
		& & ${\rm BR}(h_2\rightarrow h_1h_3)=0.86$& 13 TeV VBF:& $5.7$ fb &$d_1=6.67+i\,2.92,\,d_3=4.94-i\,10.7$\\
		& &$\Gamma_{\rm Tot}(h_2)=0.1\,m_2$& 14 TeV ggF:& $19$ fb &$e_1=(-12.0+i\,29.6)v,\,e_2=(-37.1+i\,67.6)v$\\
		& &  & 14 TeV VBF:& $6.9$ fb &$\sin\theta_1=0.201$\\ \cline{1-6}
    \end{tabular}}
\end{center}
\caption{\label{tab:bm213} Benchmark points that maximize ${\rm BR}(h_2\rightarrow h_1h_3)$ with cross sections at the LHC with $\sin\theta_1=0.201$. }
\end{table}

\begin{table}[tb]
\begin{center}
\makebox[0.9\textwidth]{
    \begin{tabular}{|c|c|c|ll|l|}\hline\hline
        $m_2$& $m_3$ & BRs and width & \multicolumn{2}{c|}{$\sigma(pp\rightarrow h_2\rightarrow h_3h_3)$} & Parameters  \\\hline\hline
       \multirow{4}{*}{400 GeV} &\multirow{4}{*}{130 GeV} & &13 TeV ggF:& $370$ fb & $d_2=18.9,\,\delta_2=1.77,\,\delta_3=-0.118+i\,0$\\
		& & ${\rm BR}(h_2\rightarrow h_3h_3)=0.97$& 13 TeV VBF:& $30$ fb &$d_1=3.14-i\,2.14,\,d_3=0.434-i\,2.62$\\
		& &$\Gamma_{\rm Tot}(h_2)=0.1\,m_2$& 14 TeV ggF:& $440$ fb &$e_1=(-8.75-i\,20.0)v,\,e_2=(-1.84+i\,60.0)v$\\
		& &  & 14 TeV VBF:& $35$ fb &$\sin\theta_1=0.201$\\ \cline{1-6}
        \multirow{12}{*}{600 GeV} &\multirow{4}{*}{130 GeV} & &13 TeV ggF:& $75$ fb & $d_2=16.5,\,\delta_2=3.12,\,\delta_3=0.604+i\,0$\\
		& & ${\rm BR}(h_2\rightarrow h_3h_3)=0.92$& 13 TeV VBF:& $12$ fb &$d_1=7.18+i\,1.47,\,d_3=-1.53-i\,6.00$\\
		& &$\Gamma_{\rm Tot}(h_2)=0.1\,m_2$& 14 TeV ggF:& $90$ fb &$e_1=(-13.0-i\,18.8)v,\,e_2=(-6.32+i\,56.5)v$\\
		& &  & 14 TeV VBF:& $15$ fb &$\sin\theta_1=0.201$\\ \cline{2-6}
          &\multirow{4}{*}{200 GeV} & &13 TeV ggF:& $75$ fb & $d_2=15.2,\,\delta_2=1.82,\,\delta_3=0.155+i\,0$\\
		& & ${\rm BR}(h_2\rightarrow h_3h_3)=0.92$& 13 TeV VBF:& $12$ fb &$d_1=1.42+i\,2.91,\,d_3=12.6+i\,5.94$\\
		& &$\Gamma_{\rm Tot}(h_2)=0.1\,m_2$& 14 TeV ggF:& $90$ fb &$e_1=(-16.9+i\,13.9)v,\,e_2=(-14.2-i\,41.7)v$\\
		& &  & 14 TeV VBF:& $15$ fb &$\sin\theta_1=0.201$\\ \cline{2-6}
        &\multirow{4}{*}{270 GeV} & &13 TeV ggF:& $75$ fb & $d_2=11.1,\,\delta_2=0.142,\,\delta_3=-0.0342-i\,0.00817$\\
		& & ${\rm BR}(h_2\rightarrow h_3h_3)=0.92$& 13 TeV VBF:& $12$ fb &$d_1=-5.13-i\,5.14,\,d_3=-3.21+i\,0.753$\\
		& &$\Gamma_{\rm Tot}(h_2)=0.1\,m_2$& 14 TeV ggF:& $90$ fb &$e_1=(-26.1-i\,12.7)v,\,e_2=(-29.7+i\,38.1)v$\\
		& & & 14 TeV VBF:& $15$ fb &$\sin\theta_1=0.201$\\ \cline{1-6}
        \multirow{12}{*}{800 GeV} &\multirow{4}{*}{130 GeV} & &13 TeV ggF:& $16$ fb & $d_2=21.1,\,\delta_2=2.42,\,\delta_3=2.42+i\,0$\\
		& & ${\rm BR}(h_2\rightarrow h_3h_3)=0.86$& 13 TeV VBF:& $5.6$ fb &$d_1=3.77-i\,8.72,\,d_3=2.21+i\,5.43$\\
		& &$\Gamma_{\rm Tot}(h_2)=0.1\,m_2$& 14 TeV ggF:& $19$ fb &$e_1=(-28.0-i\,0.44)v,\,e_2=(-41.4+i\,2.15)v$\\
		& &  & 14 TeV VBF:& $6.9$ fb &$\sin\theta_1=0.201$\\ \cline{2-6}
          &\multirow{4}{*}{200 GeV} & &13 TeV ggF:& $16$ fb & $d_2=13.8,\,\delta_2=0.810,\,\delta_3=0.810+i\,0$\\
		& & ${\rm BR}(h_2\rightarrow h_3h_3)=0.86$& 13 TeV VBF:& $5.6$ fb &$d_1=-10.8+i\,1.53,\,d_3=1.29-i\,5.41$\\
		& &$\Gamma_{\rm Tot}(h_2)=0.1\,m_2$& 14 TeV ggF:& $19$ fb &$e_1=(-32.6+i\,1.05)v,\,e_2=(-53.2-i\,8.34)v$\\
		& &  & 14 TeV VBF:& $6.9$ fb &$\sin\theta_1=0.201$\\ \cline{2-6}
        &\multirow{4}{*}{270 GeV} & &13 TeV ggF:& $16$ fb & $d_2=10.6,\,\delta_2=0.765,\,\delta_3=0.695+i\,0.145$\\
		& & ${\rm BR}(h_2\rightarrow h_3h_3)=0.86$& 13 TeV VBF:& $5.7$ fb &$d_1=0.695-i\,7.63,\,d_3=1.74-i\,4.77$\\
		& &$\Gamma_{\rm Tot}(h_2)=0.1\,m_2$& 14 TeV ggF:& $19$ fb &$e_1=(-28.3-i\,20.4)v,\,e_2=(-36.7+i\,68.7)v$\\
		& &  & 14 TeV VBF:& $6.9$ fb &$\sin\theta_1=0.201$\\ \cline{1-6}
    \end{tabular}}
\end{center}
\caption{\label{tab:bm233} Benchmark points that maximize ${\rm BR}(h_2\rightarrow h_3h_3)$ with cross sections at the LHC with $\sin\theta_1=0.201$.  }
\end{table}

In Tables~\ref{tab:bm211},~\ref{tab:bm213}, and~\ref{tab:bm233} we give the maximum branching ratios and production rates for $h_2\rightarrow h_1h_1$, $h_2\rightarrow h_1h_3$, and $h_2\rightarrow h_3h_3$, respectively, as well as the parameter points that generate these branching ratios and rates.  We choose the mass points $m_2=400,600,$ and $800$~GeV, and $m_3=130,\,200,\,$ and $270$~GeV.  The Lagrangian parameter values in these tables are not unique.  There are many possible choices that will generate the same maximum branching ratios.

When $|\sin\theta_1|\gg |\sin\theta_2|\neq0$, our approximations above is good, and $h_3$ can still decay.  If the mass of $h_3$ is below the $h_1h_1$ threshold, $h_3$ will decay like a SM Higgs with mass $m_3$.  We chose the mass points $m_3=130,\,200,$ and $270$ GeV so that $h_3$ has different decay patterns:
\begin{itemize}
\item For $m_3=130$~ the dominant decays are $h_3\rightarrow bb$ and $h_3\rightarrow WW$.  Hence, for $h_2\rightarrow h_1h_3$ and $h_2\rightarrow h_3h_3$ the dominant final states are multi-$b$ and multi-$W$.
\item For $m_3=200$~GeV, both the $WW$ and $ZZ$ thresholds open up, and by far the most dominant decay channels are $WW$ and $ZZ$.    In this case, the dominate final states for $h_2\rightarrow h_1h_3$ are $bbWW$ and $bbZZ$.  For $h_2\rightarrow h_3h_3$ the dominant final states are $4W$, $4Z$, and $WWZZ$.
\item For $m_3=270$~GeV, the $h_3\rightarrow h_1 h_1$ channel opens up.  In the small mixing limit, the relevant trilinear is
\begin{eqnarray}
h_1h_1h_3: -{\rm Im}(\delta_3)\,v\,\sin\theta_1+\mathcal{O}(\sin^2\theta_1,\sin\theta_2)
\end{eqnarray}
hence, the branching ratio of $h_3\rightarrow h_1h_1$ can be substantial.   Hence, it is possible to have a dominant signature be cascade Higgs decays: $h_2\rightarrow h_1h_3\rightarrow 3\,h_1$ and $h_2\rightarrow h_3h_3\rightarrow 4\,h_1$.  
\end{itemize}

\section{Conclusion}
\label{sec:Conclusion}
Extended scalar sectors are a feature of many models. Scalar singlets are a simple, but phenomenologically interesting, way to extend the Standard Model. The complex singlet extension, in particular, allows for resonant production of multiple different two scalar final states. In this work, we found benchmarks for resonant production and decays $pp\rightarrow h_2 \rightarrow h_1 h_1$,  $pp\rightarrow h_2 \rightarrow h_1 h_3$, and $pp\rightarrow h_2 \rightarrow h_1 h_3$ in the complex singlet model.

For a variety of masses, we consistently find that the branching ratios for $h_2 \rightarrow h_i h_j$ can consistently be around $0.7-1$. This demonstrates the importance of double Higgs searches, particularly those where the final state ``Higgs bosons'' could be scalars other than the Standard Model-like Higgs boson. The typical ``Higgs-like'' decays of scalars to Standard Model fermion and gauge boson final states for $h_2$ are subdominant for these benchmarks. Additionally, the decays of $h_2$ is the main production mode of $h_3$ in the limit of small mixing, since all the couplings of $h_3$ to Standard Model fermions and gauge bosons are double mixing angle suppressed.  For the complex singlet benchmarks we have presented, these generalized double Higgs channels are the essential discovery channels.

\section*{Acknowledgements}
SA, SDL, IML, and MS have been supported in part by the United States Department of Energy
grant number DE-SC001798.  SA, SDL,  MS are also supported in part by
the State of Kansas EPSCoR grant program. MS is also supported in part by the United States Department of Energy under Grant Contract DE-SC0012704.  SDL was supported in part by the University of Kansas General Research Funds.  Data for the plots is available upon request.

\bibliographystyle{utphys}
\bibliography{references}

\end{document}